# 2D material platform for overcoming the amplitude-phase tradeoff in ring modulators

Ipshita Datta[1], Andres Gil-Molina[1], Sang Hoon Chae[2,3], James Hone[2] and Michal Lipson[1]

[1]Department of Electrical Engineering, Columbia University, New York, New York 10027, USA

[2]Department of Mechanical Engineering, Columbia University, New York, New York 10027, USA

[3]School of Electrical & Electronic Engineering, School of Materials Science and Engineering, Nanyang Technological University, Singapore 639798, Singapore

Corresponding Author – ml3745@columbia.edu

**Abstract –** Compact, high-speed electro-optic phase modulators play a vital role in various large-scale applications including phased arrays, quantum and neural networks, and optical communication links. Conventional phase modulators suffer from a fundamental tradeoff between device length and optical loss that limits their scaling capabilities. High-finesse ring resonators have been traditionally used as compact intensity modulators, but their use for phase modulation have been limited due to the high insertion loss associated with the phase change. Here, we show that high-finesse resonators can achieve a strong phase change with low insertion loss by simultaneous modulation of the real and imaginary parts of the refractive index, to the same extent. To implement this strategy, we utilize a hybrid platform that combines a low-loss SiN ring resonator with electro-absorptive graphene (Gr) and electro-refractive $WSe_2$. We achieve a phase modulation efficiency ($V_{\pi/2} \cdot L_{\pi/2}$) of 0.045 V · cm with an insertion loss ($IL_{\pi/2}$) of 4.7 dB for a phase change of $\pi/2$ radians, in a 25 µm long $Gr-Al_2O_3-WSe_2$ capacitor embedded on a SiN ring of 50 µm radius. We find that our $Gr-Al_2O_3-WSe_2$ capacitor can support an electro-optic bandwidth of 14.9 ± 0.1 GHz. We further show that the $V_{\pi/2} \cdot L_{\pi/2}$ of our SiN-2D platform is at least an order of magnitude lower than that of electro-optic phase modulators based on silicon, III-V on silicon, graphene on silicon and lithium niobate. This SiN-2D hybrid platform provides the impetus to design compact and high-speed reconfigurable circuits with graphene and transition metal dichalcogenide (TMD) monolayers that can enable large-scale photonic systems.

# Introduction

Conventional photonic materials used in phase modulators, typically exhibit low index change ranging from 0.01 % in lithium niobate (LiNbO$_3$) to 0.1 % for semiconductors such as silicon and III-V (see supplementary section I). This low index change requires hundreds of microns of interaction length for a phase change of $\pi$ radians[1–25]. In high-speed intensity modulators[26,16,27,28], it is possible to achieve effective modulation in a compact form-factor by using high-finesse ring resonators to increase the interaction length and enhance modulation depth.

However, high-finesse ring resonators have not been successfully implemented for phase modulation, due to the associated large insertion loss and undesired intensity modulation. This loss arises due to the effect of a large mismatch between the change in real ($\Delta n$) and imaginary part ($\Delta k$) of the index for conventional materials used for phase modulation in ring resonators. As an example, in figure 1a, we show the transmission ($T_{Ring}$) and phase response ($\varphi_T$) of a ring resonator embedded with an active material that undergoes a strong change in index i.e. $\frac{\Delta n}{\Delta k} \sim -20$, typically observed in silicon. The ring is critically-coupled with a loaded quality factor ($Q_L$) of ~ 20,000 and undergoes a $\Delta n$ of -6 × 10$^{-4}$ RIU (refractive index units), as is commonly reported in silicon-based electro-refractive modulators[28–30]. The index change in the active material induces a shift in the resonance wavelength of the ring response. At a probe wavelength of $\lambda_p$, the cavity experiences a phase change of ~ $\pi$/2 radians that is accompanied by an unacceptably high insertion loss of 9 dB. Previous demonstrations have leveraged low-finesse over-coupled ring resonators for phase modulation[31,32]. However, these modulators require high index change and therefore, must rely on the thermo-optic effect which is an inherently slow and power-hungry mechanism.

Here, we demonstrate an alternative strategy for achieving strong phase change in resonators with low insertion loss and minimal transmission variation. This strategy requires that the ratio of the index change ($\Delta n$) to the loss change ($\Delta k$) has to be approximately equal to unity, i.e. $\frac{\Delta n}{\Delta k} \sim 1$ (see Supplementary section II). In figure 1b, we show the $T_{Ring}$ and $\varphi_T$ at the output of a ring resonator utilizing an embedded hybrid active material with $\frac{\Delta n}{\Delta k} \sim 1$, operating near the critically-coupled regime with an initial $Q_L$ of ~ 20000 and a $\Delta n$ of -6 × 10$^{-4}$ RIU. In this case, $\Delta n$ causes resonance detuning, while $\Delta k$ results in an enhancement of the coupling between the ring resonator and the bus waveguide. One can see from the phase profile in figure 1b, that these effects transform the $\varphi_T$ from a gradual phase variation near the critically-coupled condition to a strongly dispersive phase profile in the over-coupled regime with a detuning of the resonance wavelength. When probed at $\lambda_p$, one can achieve a strong phase change of ~ π radians with an insertion loss as low as 4.5 dB. However, no known dielectric material exhibits a $\frac{\Delta n}{\Delta k} \sim 1$ at near-infrared (NIR) wavelengths.

To enable simultaneous tuning of both the real and imaginary part of the effective index, we utilize a capacitive stack of monolayer tungsten disulphide (WSe$_2$) and graphene (Gr) integrated in a low-loss dielectric ring resonator. Figure 1c shows a schematic of the structure consisting of the stack of the 2D materials (Gr-Al$_2$O$_3$-WSe$_2$) embedded on a 1300 nm wide × 330 nm high silicon nitride (SiN) waveguide covered with 180 ± 15 nm of planarized silicon dioxide (SiO$_2$). We tune the electro-optic properties of the 2D materials by applying a voltage across the dielectric alumina (Al$_2$O$_3$) that is sandwiched between the two monolayers. Our design benefits from the electro-absorptive properties of Gr[34–36] (which effectively tunes the ring-bus coupling) and the electro-refractive properties of monolayer transition metal-dichalcogenide (TMD) such as WSe$_2$[37] (which changes the resonance wavelength). The ratio between the real ($\Delta n_{\text{eff}}$) and imaginary part ($\Delta k_{\text{eff}}$) of the effective index in a composite SiN-2D waveguide, is determined by the relative overlap of the propagating mode with each of the 2D monolayers. The overlap with Gr influences both the initial propagation loss and $\Delta k_{\text{eff}}$, while the overlap with the monolayer WSe$_2$ predominantly contributes to the $\Delta n_{\text{eff}}$. The bare SiN resonator with a radius of 50 µm, is fabricated with an intrinsic quality factor ($Q_0$) of ~ 112,000 (i.e. $\alpha_{\text{SiN}}$ ~ 5.98 ± 0.38 dB/cm). The ring resonator is designed to achieve critical-coupling with a $Q_L$ of ~ 20,870 at 8 V, by embedding a 40 µm long Gr-Al$_2$O$_3$-WSe$_2$ capacitor in the ring with a ring-bus gap of 350 nm (hereafter, referred to as device I). The SiO$_2$-clad SiN waveguides were fabricated using standard techniques, and then planarized to permit the mechanical transfer and lithographic patterning of the Gr-Al$_2$O$_3$-WSe$_2$ capacitor with metal contacts in the SiN ring resonator. We estimate a linear induced charge density of (0.85 ± 0.03) × 10$^{12}$ cm$^{-2}$ per volt on both the monolayers, based on the thickness (45 nm) and the extracted dielectric permittivity of Al$_2$O$_3$ (6.9 ± 0.2) (see Methods).

## Results

We measure a continuous phase change ($\Delta \varphi_T$) of (0.46 ± 0.05) π radians with an insertion loss (IL) of 4.78 ± 0.40 dB. This $\Delta \varphi_T$ is accompanied with a transmission variation ($\Delta T_{Ring}$) of 4.37 ± 0.70 dB for an applied voltage swing from 6 V to 18 V. In figure 2a, we show using blue square markers, the $\Delta \varphi_T$ (top panel) and the $\Delta T_{Ring}$ (bottom panel) for different voltages applied across the Gr-Al$_2$O$_3$-WSe$_2$ capacitor. We probe the $\Delta \varphi_T$ and $\Delta T_{Ring}$ at a wavelength detuning $\lambda_p$, where the phase change is maximum while ensuring a low IL and $\Delta T_{Ring}$. In this case, $\lambda_p$ is 0.03 nm blue-detuned from the resonance at critical coupling (1538.71 nm). The $\Delta T_{Ring}$ for different voltages is normalized with respect to the IL, measured at an initial bias voltage of 6 V. One can confirm from the $T_{Ring}$ spectra of device I in the bottom right inset of figure 2a, that the ring is initially in the under-coupled regime at 0 V. As we increase the applied voltage to 8 V, the resonator becomes critically-coupled and at voltages exceeding 10 V, the ring becomes over-coupled with a strong blue detuning of the resonance wavelength (for detailed results, see Supplementary section III).

We measure a 3 dB electro-optic bandwidth of 14.9 ± 0.1 GHz in the composite SiN-2D waveguide. In figure 2b, we show the frequency response of device I, which we measure using a 70 GHz fast photodiode and an electrical vector network analyzer (VNA) at 1569.6 nm for a bias voltage of 8 V with a RF voltage swing of 10 dBm. One can see from the measured $T_{Ring}$ at 1569.6 nm in the inset of figure 2b, that the optical bandwidth supported by the device at 8 V is 15.8 GHz ($Q_L \sim 12000$). We ensure that the optical bandwidth supported by the device at 8 V is higher than the measured electro-optic bandwidth (14.9 GHz) that is currently limited by the contact resistance of monolayer Gr and $WSe_2$ (see supplementary section IV).

We show that the voltage dependent phase and transmission change in ring resonators can be tailored with the device geometry. We engineer the length of $Gr-Al_2O_3-WSe_2$ capacitor to achieve a similar phase change of (0.50 ± 0.05) π radians as observed in device I, while ensuring a comparatively lower IL of 2.96 ± 0.34 dB and low $\Delta T_{Ring}$ of 1.73 ± 0.20 dB. The optimized device (hereafter, referred to as device II) consists of a 25 µm-long $Gr-Al_2O_3-WSe_2$ capacitor embedded in the SiN ring, with a ring-bus gap of 450 nm. The shorter capacitor exhibits lower pin-hole defects that results in an increase of the breakdown voltage from 22 V in device I to 30 V in device II. The high breakdown fields enable a higher degree of transparency in SiN waveguide, thereby facilitating strong tuning of the coupling regime (see Supplementary section V). Similar to the configuration of device I, the ring achieves critical coupling with a $Q_L$ of ~ 18,730 at 8.5 V. We probe the $\Delta\varphi_T$ (top panel of figure 2a) and $\Delta T_{Ring}$ (bottom panel of figure 2a) at a $\lambda_p$, that is blue detuned by 0.04 nm with respect to the resonance wavelength at critical coupling (1646.22 nm). The $\Delta\varphi_T$ of the ring are extracted from the measured $T_{Ring}$ for different voltages applied across the capacitor (see Supplementary section VI). We further confirm that the $\Delta\varphi_T$ extracted from the $T_{Ring}$ is in strong agreement with the $\Delta\varphi_T$ measured using an external Mach-Zehnder interferometer (MZI) configuration (often used for phase measurements)[38,39,31]. We measure $\Delta\varphi_T$ by simultaneously probing both the $T_{Ring}$, and the transmission at the output of an MZI ($T_{MZI}$) when device I is embedded in one of the arms of a fiber-based MZI. Supplementary section VII details the experimental setup used for measuring the phase response of our devices, while supplementary section VIII shows the measured $T_{Ring}$ and $T_{MZI}$ at the output of the ring resonator, for different voltages applied across the Gr-Al$_2$O$_3$-WSe$_2$ capacitor.

We verify that the ratio between the $\Delta n_{eff}$ and $\Delta k_{eff}$ of the composite SiN-2D waveguide is close to unity. Figure 3a shows the measured $\Delta n_{eff}$ and $\Delta k_{eff}$ of the ring, which is extracted from the measured $T_{Ring}$ for various voltages applied across the 25 μm long Gr-Al$_2$O$_3$-WSe$_2$ capacitor in device II (see Supplementary section IX). One can see that the $\Delta k_{eff}$ decreases progressively with an increase in the gate voltage and $\Delta n_{eff}$ shows an initial increase leading to a red-shift in the resonance wavelength, followed by a strong decrease which leads to a strong blue detuning. The maximum $\Delta n_{eff}$ and $\Delta k_{eff}$ for an applied voltage of 30 V is -6.0 × 10$^{-4}$ RIU and -7.8 × 10$^{-4}$ RIU respectively, which corresponds to a $\frac{\Delta n_{eff}}{\Delta k_{eff}}$ ~ 0.78.

We extract the contribution of the electro-optic response of each of the 2D layers to the $\Delta n_{eff}$ and $\Delta k_{eff}$, by modeling the monolayers as a hybrid 2D sheet integrated on a SiN waveguide using finite element model (see Methods and Supplementary Section X). Figure 3b shows the change in the normalized imaginary and real part of the optical conductivity of monolayer graphene $\left(\frac{\sigma_G}{\sigma_0}\right)$ that imparts a proportional change in the $\Delta n_{eff}$ and $\Delta k_{eff}$ of the propagating mode, respectively. From the modeling of the Gr monolayer, we find that the Gr is initially p-doped with $(5.20 \pm 0.30) \times 10^{12}$ cm$^{-2}$ carriers (i.e. $E_{F_{init}}= 0.240 \pm 0.006$ eV) and becomes completely transparent at voltages exceeding 25 V. Figure 3c shows the change in the real and imaginary part of the refractive index of WSe$_2$ ($\Delta n_{WSe_2}$ and $\Delta k_{WSe_2}$) as a function of the applied voltage. For a maximum electron doping of $(2.54 \pm 0.74) \times 10^{13}$ cm$^{-2}$ at 30 V, the $\Delta n_{WSe_2}$ reaches $-0.69 \pm 0.05$ RIU, while $\Delta k_{WSe_2}$ (which induces insertion loss) remains below $0.010 \pm 0.002$ RIU.

We measure a phase modulation efficiency ($V_{\pi/2} \cdot L_{\pi/2}$) of 0.045 V · cm with an insertion loss (IL$_{\pi/2}$) of 4.7 dB for a phase change of $\pi/2$ radians. We show in figure 4 and Supplementary section XI, that device II performs with a significantly lower $V_{\pi/2} \cdot L_{\pi/2}$ (i.e., enhanced phase modulation efficiency) and lower IL$_{\pi/2}$, when compared to the existing phase modulation technologies. We find that the $V_{\pi/2} \cdot L_{\pi/2}$ for our SiN-2D hybrid platform is at least an order of magnitude lower than the one of electro-refractive phase modulators based on dielectric materials such as silicon PN, PIN and MOS capacitors[1,3,7,9–11], III-V on silicon[22], graphene on silicon modulators[18] and lithium niobate (LN) devices[16,17], and can be achieved with a relatively low IL$_{\pi/2}$. In figure 4, we show a comparison with current state-of-art phase modulation technologies. One can see that by overcoming the traditional tradeoff between propagation length and insertion loss through simultaneous index and loss modulation in ring resonators, our SiN-2D platform facilitates the development of compact phase modulators with low $V_{\pi/2} \cdot L_{\pi/2}$ and minimal optical loss.

The SiN-2D hybrid platform enables the design of compact and highly reconfigurable photonic circuits with tunable coupling and the ability to achieve phase modulation at several gigahertz of electro-optic bandwidth. We show a novel paradigm of designing efficient and compact phase modulators by leveraging cavities embedded with a hybrid material that has a $\frac{\Delta n}{\Delta k}$ close to unity. Alternately, one can realize high-speed intensity modulation with low insertion loss using the same SiN-2D platform, by probing the response at the resonance wavelength[40,41]. The potential of our SiN-2D platform to become transparent with doping and the ability to modify the coupling in cavities at several gigahertz of electro-optic bandwidth enables its use in various applications such as optical memories, frequency combs and optical communication systems[41].

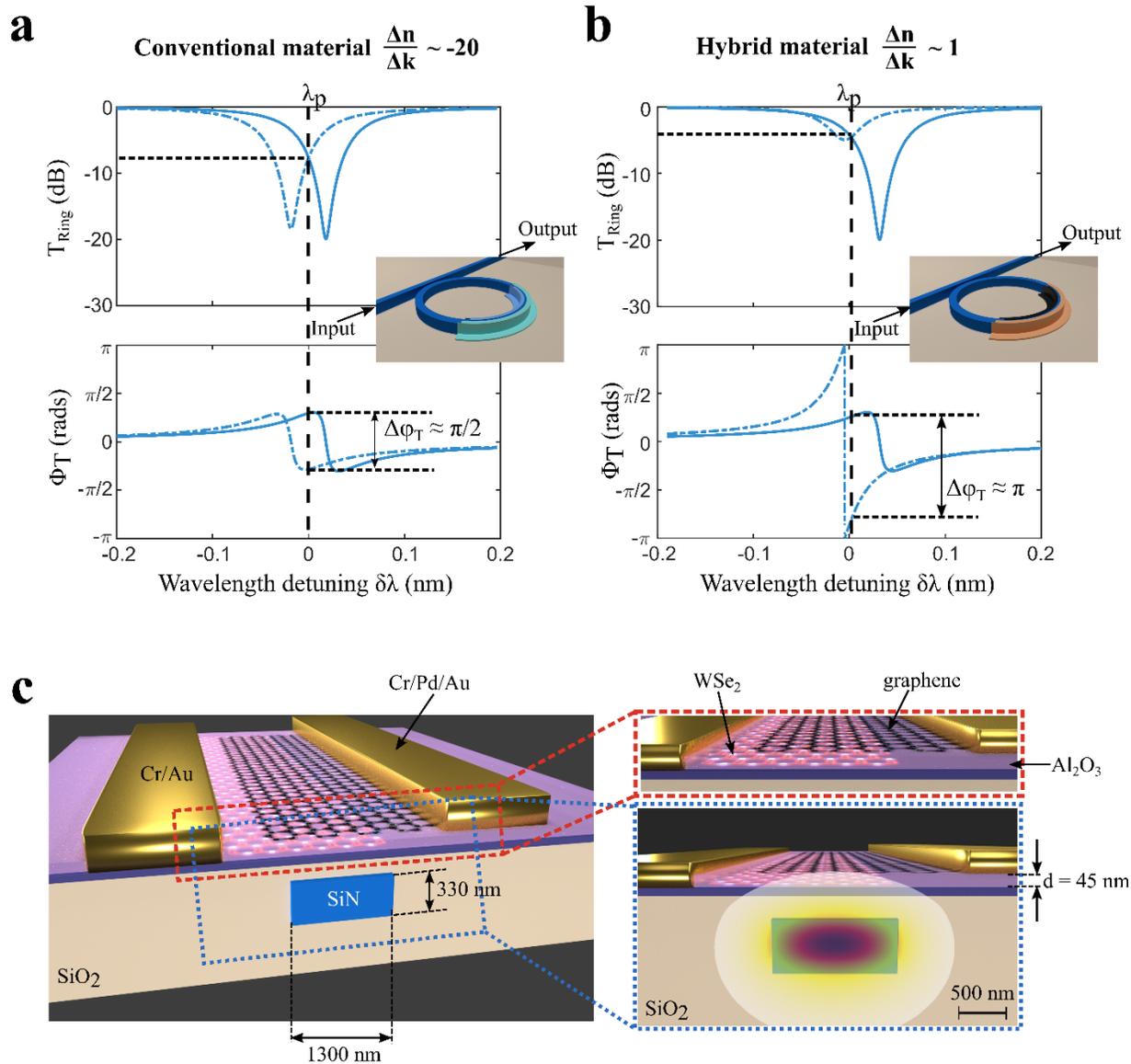

**Figure 1 | Illustration of SiN-2D platform leveraging loss and index change for phase modulation.** (a) Normalized transmission ($T_{Ring}$) and phase response ($\varphi_T$) of a ring resonator embedded with a conventional material that undergoes a strong change in index i.e. $\frac{\Delta n}{\Delta k} \sim -20$. The inset shows the schematic of a ring resonator with a portion (~ 8 %) of the ring covered with the active conventional material. For a ring operating near the critically-coupled regime with a loaded quality factor ($Q_L$) of ~ 20000, the $\Delta n$ of -6 × 10⁻⁴ RIU induces a blue shift in the resonance wavelength of the ring response. At a probe wavelength of $\lambda_p$, the cavity experiences a strong phase change that is accompanied with a prohibitively high insertion loss. (b) $T_{Ring}$ and $\varphi_T$ of a ring resonator embedded with a hybrid material that undergoes simultaneous index and loss change i.e. $\frac{\Delta n}{\Delta k} \sim 1$. For a similar ring with ~ 8 % of the ring

covered with the hybrid material, with the ring $Q_L$ of ~ 20000 and a $\Delta n$ of -6 × 10⁻⁴ RIU, one can observe a strong blue shift in the resonance wavelength with drastic change in the coupling regime. When probed at $\lambda_p$, one can achieve a strong phase change at of ~ π radians with an insertion loss as low as 4.5 dB. (c) Illustration of a composite SiN-2D hybrid platform with monolayer Gr and $WSe_2$ to tune the loss and index of SiN waveguide, respectively. Device cross-section showing Gr-$Al_2O_3$-$WSe_2$ capacitor, embedded on a 1300 nm wide × 330 nm high SiN waveguide. Top right inset shows the Gr-$Al_2O_3$-$WSe_2$ parallel plate capacitor configuration which electrostatically gates both the monolayers by applying a voltage across the dielectric. Bottom right inset indicates the position of the monolayer Gr and $WSe_2$ in the mode-profile of the SiN waveguide.

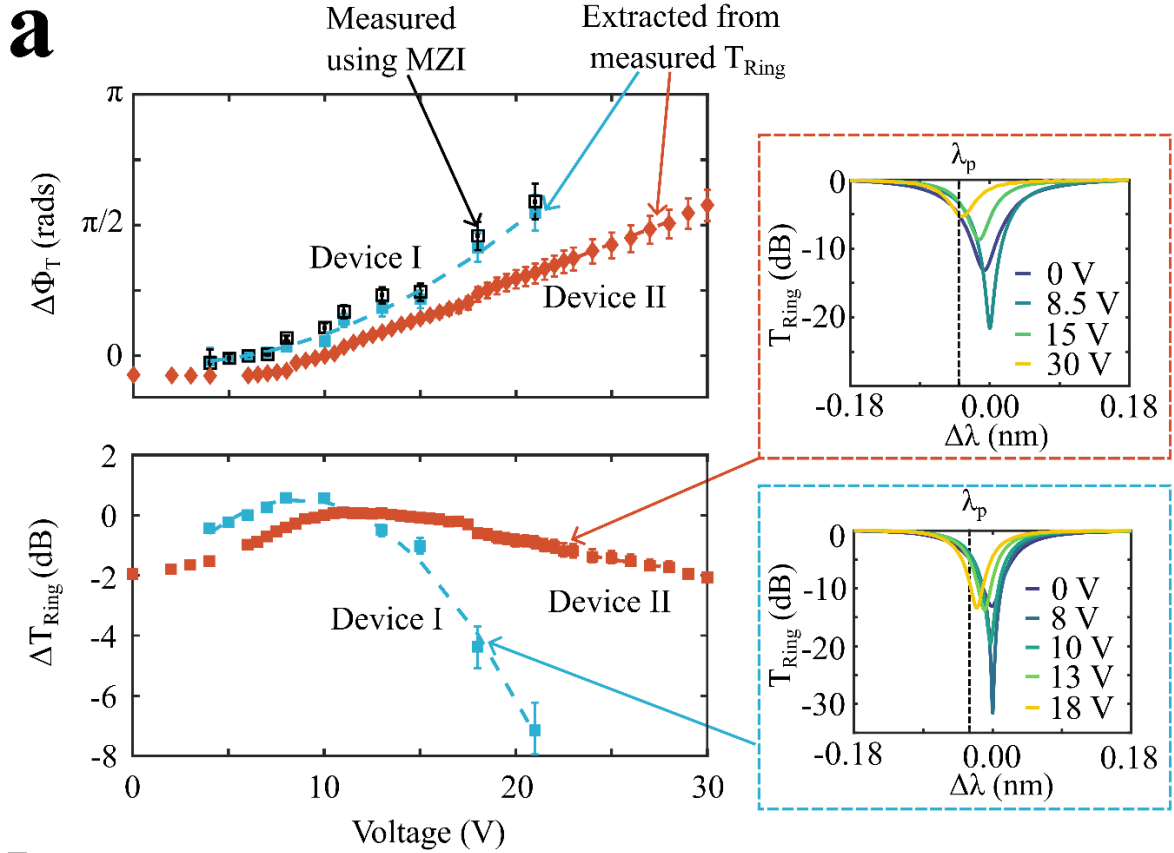
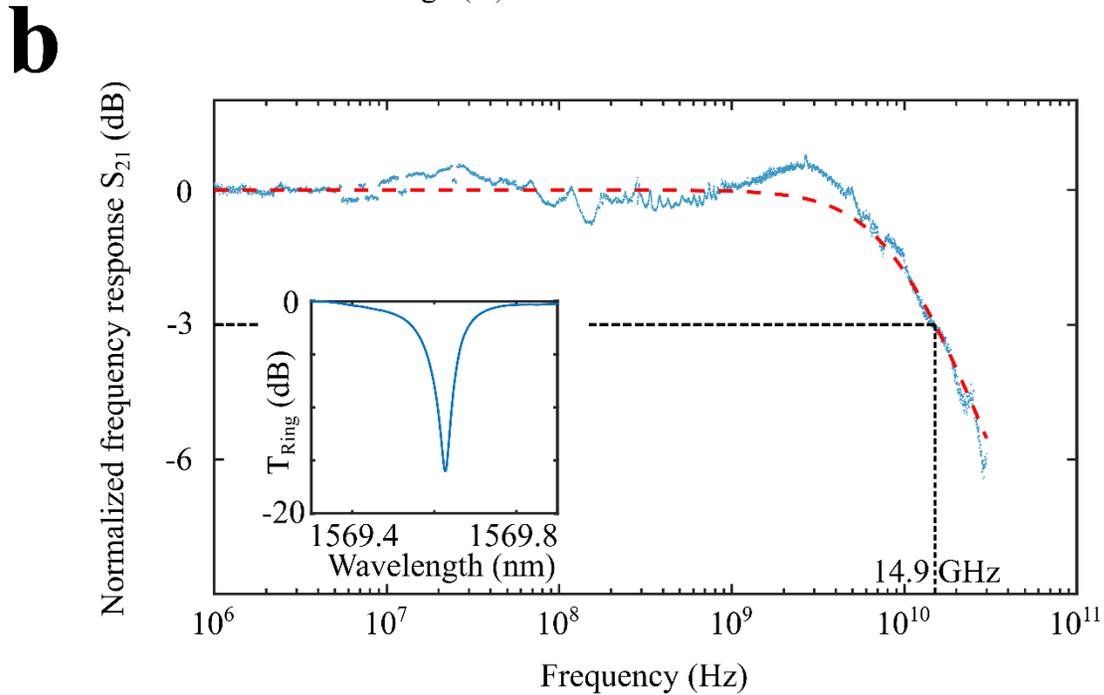

**Figure 2: Phase measurement and electro-optic bandwidth of SiN-2D hybrid waveguide embedded in a SiN ring resonator.** (a) Phase change ($\Delta\varphi_T$ in radians) in the top panel and transmission modulation ($\Delta T_{Ring}$ in dB) in the

bottom panel at probe wavelength detuning ($\lambda_p$), for different voltages applied across two Gr-Al$_2$O$_3$-WSe$_2$ capacitors embedded in a SiN ring resonator of 50 µm radius. The blue markers show the extracted $\Delta\varphi_T$ and measured $\Delta T_{Ring}$ for device I. One can see a continuous phase change ($\Delta\varphi_T$) of (0.46 ± 0.05) π radians in device I with an insertion loss (IL) of 4.78 ± 0.40 dB and $\Delta T_{Ring}$ of 4.37 ± 0.70 dB for an applied voltage swing from 6 V to 18 V. The orange markers show the extracted $\Delta\varphi_T$ and measured $\Delta T_{Ring}$ for device II. One can see that the device II can enable a $\Delta\varphi_T$ of (0.50 ± 0.05) π radians with a comparatively lower IL of 2.96 ± 0.34 dB and lower $\Delta T_{Ring}$ of 1.73 ± 0.20 dB. The $T_{Ring}$ spectra for device I and II shows that the rings are initially in the under-coupled regime at 0 V, becomes critically-coupled at 8.5 V and enters the over-coupled regime at voltages exceeding 10 V. We further confirm using device I, that the $\Delta\varphi_T$ extracted from the $T_{Ring}$ correlates strongly with the $\Delta\varphi_T$ measured using a Mach-Zehnder interferometer (MZI) configuration (see black markers). (b) Normalized frequency response ($S_{21}$) of device I at 1569.6 nm for a bias voltage at 8 V.

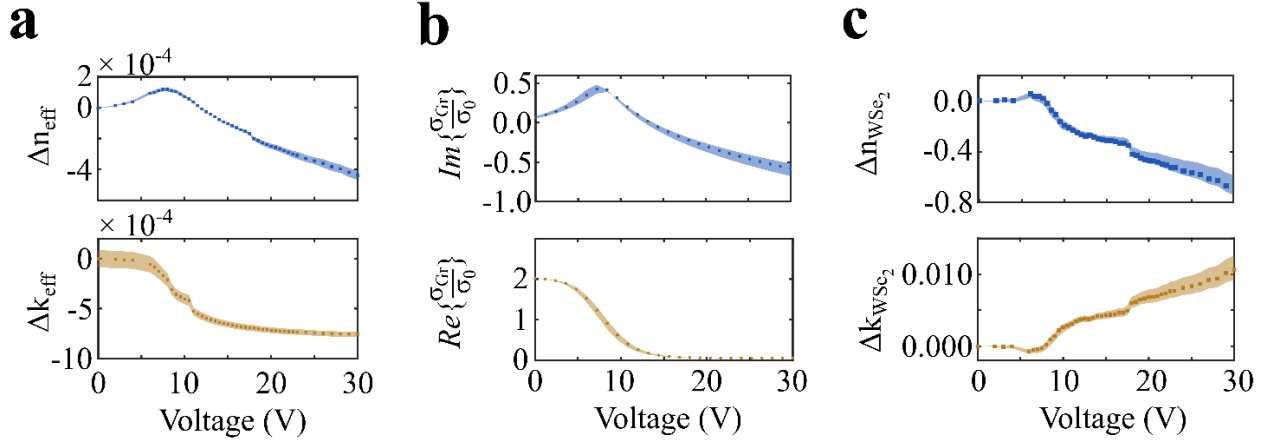

**Figure 3: Change in complex effective index of the mode and relative contribution of the Gr and WSe$_2$ monolayers.** (a) Change in the real (top) and imaginary (bottom) part of the effective index ($\Delta n_{eff}$ and $\Delta k_{eff}$ in refractive index units (RIU)) of the composite SiN-2D waveguide at different voltages, extracted from the normalized $T_{Ring}$ of device II. The shaded area represents the root-mean-square (r.m.s.) error from the numerical fit[33] (see Supplementary section X). (b) Change in the imaginary (top) and real (bottom) part of the normalized optical conductivity of monolayer graphene $\left(\frac{\sigma_G}{\sigma_0}\right)$ that imparts a proportional change in the $\Delta n_{eff}$ and $\Delta k_{eff}$ of the propagating mode, respectively. The shaded area includes the r.m.s error in the effective index and the error in the extracted initial doping of monolayer graphene, dielectric permittivity of $Al_2O_3$ and the variation in the height of the $SiO_2$ cladding separating the capacitive stack from the SiN waveguide (180 ± 10 nm). (c) Change in the real (top) and imaginary (bottom) part of the refractive index of monolayer WSe$_2$ ($\Delta n_{WSe_2}$ and $\Delta k_{WSe_2}$ in RIU) with voltage. The shaded region incorporates the r.m.s error in the complex effective index, error in the $\left(\frac{\sigma_G}{\sigma_0}\right)$ of graphene and a ± 0.05 nm variation in the thickness of the monolayer WSe$_2$ ($h_{WSe_2}$ ~ 0.65 nm)[42,43].

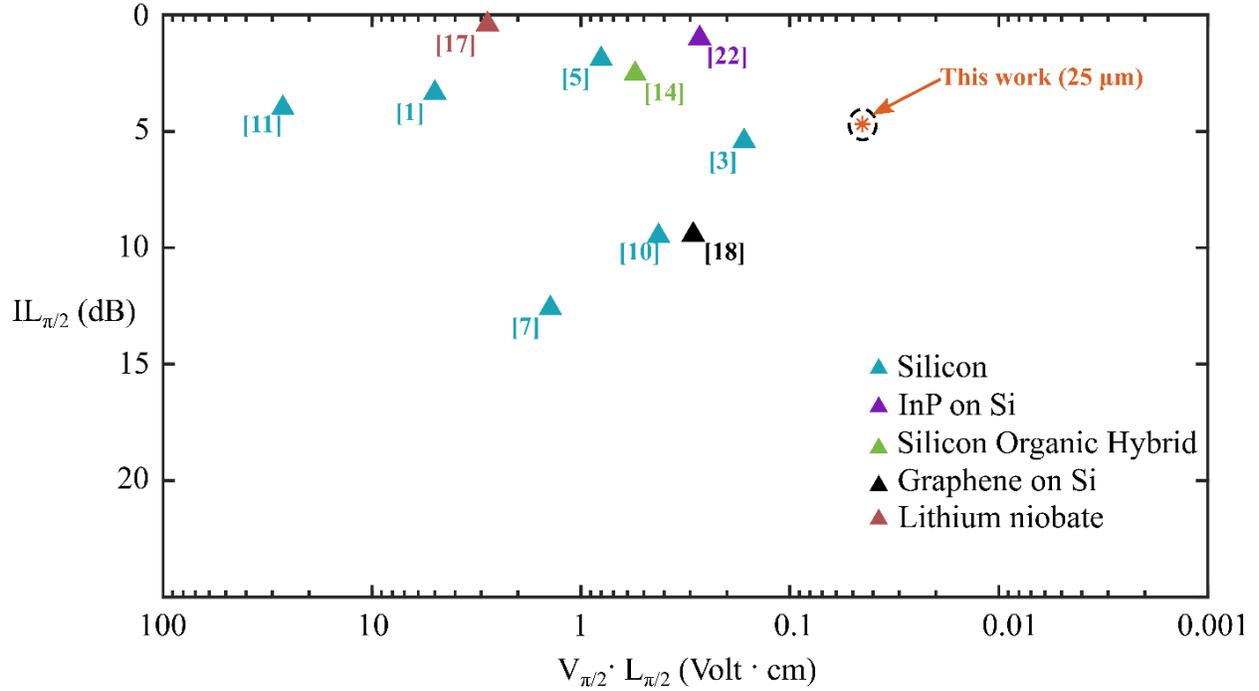

**Figure 4: Comparison of insertion loss ($IL_{\pi/2}$) vs. phase modulation efficiency (i.e. voltage length product ($V_{\pi/2} \cdot L_{\pi/2}$)) for various electro-optic phase modulators.** We show that the $V_{\pi/2} \cdot L_{\pi/2}$ of our device II (0.045 V · cm) is atleast an order of magnitude lower than conventional electro-refractive phase modulators based on silicon[1,3,5,7,10,11], III-V materials such as InP on silicon [22], graphene on silicon[18], slot-based silicon-organic hybrid material[14]. Low-loss phase modulation can be achieved by using LN devices[16,17], which comes at the expense of a $V_{\pi/2} \cdot L_{\pi/2}$ that is almost two-orders of magnitude higher than our device.

## Acknowledgement

Research on tunable electro-optic phenomenon in TMD semiconductors and graphene monolayers was supported as part of Energy Frontier Research Center (EFRC) on Programmable Quantum Materials funded by the U.S. Department of Energy (DOE), Office of Science, Basic Energy Sciences (BES), under award #DE-SC0019443. Research of I.D. on leveraging 2D materials for low-loss phase shifters was funded by Defense Advance Projects Agency (DARPA) award no. #FA8650-16-7643 and the Air Force Office of Scientific Research (AFOSR) MURI award no. N00014-16-1-2219. The platform development was partially supported by the Co-design Center for Quantum Advantage (C2QA – BNL No.390033). This work was done in part at the City University of New York Advanced Science Research Facility. We acknowledge the use of facilities and instrumentation supported by NSF through the Columbia University, Columbia Nano Initiative, and the Materials Research Science and Engineering Center DMR-2011738. We thank Dr. Brian S. Lee for his guidance with the improved graphene transfer and Dr. Gaurang R. Bhatt, Dr. Janderson Rocha Rodrigues, Dr. Utsav Deepak Dave and Dr. Aseema Mohanty for fruitful discussions.

## Contributions

I.D. and M.L. conceived and proposed the SiN-2D photonic design and experiments. I. D. fabricated the composite SiN-2D photonic device with assistance from S. H. C. for TMD transfer. I. D. performed and analyzed the optical experiments, with assistance from A. G. M. regarding the external fiber-based MZI setup for measuring phase in ring resonators. I. D. and M. L. prepared the manuscript. A. G. M., S. H. C., J. H. and M. L. edited the manuscript. M. L. supervised the project.

## Methods

### WSe$_2$ transfer and patterning

We leverage the facile method described in Ref [44] to exfoliate large-area monolayer WSe$_2$ onto our SiN waveguides, covered with 180 nm of planarized SiO$_2$. We start with an atomically flat gold film, deposited by evaporating 150 nm thin Au films onto an ultra-flat surface of highly polished silicon wafer, where the gold film is stripped away off the substrate using a combination of the thermal release tape with a polyvinylpyrrolidone (PVP) interfacial layer. The ultra-flat gold tape allows for a uniform contact between the gold and monolayer WSe$_2$ crystal surface (HQ Graphene- http://www.hqgraphene.com/WSe2.php), exfoliating a complete monolayer that can be transferred onto our planarized SiN waveguides. We remove the thermal release tape by heating our substrate to 100 °C, washing off the PVP layer and etching the gold with a mild solution of gold etchant (I$_2$/I$^-$). Supplementary figure S21 shows the extent of the coverage of monolayer WSe$_2$ after the gold assisted transfer of WSe$_2$ onto our planarized SiN substrates. Supplementary figure S22 shows the photoluminescence (PL) spectrum of as transferred monolayer WSe$_2$ on our SiO$_2$ covered SiN substrate using a Renishaw InVia Micro-Raman spectrometer at an excitation wavelength of 532 nm.

We pattern a 50 µm long WSe$_2$ monolayer by spinning a dual resist mask of 400 nm/120 nm PMMA/HSQ (XR-1561 6%) film, followed by baking the pattern at 180 °C for 15 mins (PMMA)/4 mins (HSQ), respectively, patterning using EBL and reactive ion etching (RIE) based O$_2$ plasma treatment for 4 min 30 secs to etch the residual PMMA and monolayer WSe$_2$. After the etch, we strip the resist in acetone, where it dissolves the PMMA, cleanly removing the HSQ mask.

## Graphene transfer and patterning

We use chemical vapor deposited (CVD) graphene grown on 3-inch × 3-inch copper films (Grolltex - https://grolltex.com/). We prepare the graphene samples for transfer by first spinning PMMA 495 A6 at 1000 rpm and drying the 500 nm PMMA coated graphene on Cu film overnight in ambient conditions. We electrochemically delaminate the PMMA/graphene stack from the Cu film using the process described in Ref [45]. We prepare 1M NaOH aqueous solution as an electrolyte and delaminate the PMMA/Gr stack by using the PMMA/Gr on Cu foil as the cathode, and a bare Cu foil as the anode. The delaminated PMMA/Gr stack is then transferred to a fresh water bath and this process is repeated a few times, before being transferred onto the SiN substrate. We enhance the hydrophilicity of the substrate and remove moisture/polymer contamination by performing O$_2$ plasma clean on the sample for 30 minutes prior to the transfer. Following the transfer, we vacuum dry the as transferred sample overnight in a vacuum desiccator, followed by baking the sample at 180 °C for 2 hours. Finally, the PMMA is dissolved away in acetone solution by submerging the chip in acetone for about 4 hours. Supplementary figure S23 shows the Raman spectra of the top graphene sheet after the transfer.

We pattern a 25 µm / 40 µm long graphene monolayer by spinning a composite resist mask of 400 nm/120 nm PMMA/HSQ (XR-1561 6%) film, followed by baking the pattern at 180 °C for 15 mins (PMMA)/4 mins (HSQ), respectively, patterning using EBL and reactive ion etching (RIE) based O$_2$ plasma treatment for 1 min 30 secs to etch the residual PMMA and graphene. After the etch, we strip the resist in acetone, where it dissolves the PMMA, cleanly removing the HSQ mask.

## Device fabrication for Gr-Al$_2$O$_3$-WSe$_2$ based capacitive SiN photonic device

We lithographically defined 1.3 µm wide waveguides on 330 nm high silicon nitride (SiN), deposited using Low Pressure Chemical Vapor Deposition (LPCVD) at 800 °C and annealed at 1200 °C for 3 hours on 4.2 µm thermally oxidized SiO$_2$, using a combination of deep ultraviolet (DUV) lithography to define the chemical planarization (CMP) pillars of 5 µm length × 5 µm width, with 33 % fill factor in the wafer area, surrounding the waveguides and ebeam lithography (EBL) to define the waveguides. In order to obtain low-loss SiN waveguides at near infrared (NIR) wavelengths, we leverage an optimized etch recipe, described in Ref [46] to reduce the surface roughness of SiN waveguides that contributes to the propagation loss in low confinement SiN waveguides. We etch the SiN waveguides and CMP patterns using an optimized $CHF_3/O_2$ recipe with increased oxygen flow to reduce *in situ* polymer formation in Oxford 100 Plasma ICP RIE, using 360 nm of PECVD $SiO_2$ as a hard mask for etching the SiN thin film. We remove the residual $SiO_2$ hard mask using a 100:1 buffered oxide etch solution (BOE) to reduce the roughness due to etch, followed by deposition of 600 nm of Plasma Enhanced Chemical Vapor Deposition (PECVD) silicon dioxide ($SiO_2$) on the waveguides for planarization. We planarize the $SiO_2$ to 180 nm ± 15 nm above the SiN waveguides using standard CMP techniques to create a planar surface for the transfer of monolayer TMD such as $WSe_2$ and to prevent the $WSe_2$ film from breaking at the waveguide edges. We clean the planarized surface with Piranha solution at 100 C to remove the slurry particles that settle during CMP process. The 180 nm $SiO_2$ layer additionally aids in reducing the optical propagation loss introduced by the interaction of the undoped graphene monolayer with the optical mode. A 15 nm of sacrificial thermal atomic layer-deposited (ALD) alumina ($Al_2O_3$) is deposited on top of $SiO_2$ to isolate the SiN waveguides from the subsequent fabrication steps required for the patterning of monolayer TMDs. Following the $WSe_2$ transfer and patterning steps described above, the metal contacts are lithographically patterned using EBL, and 0.5 nm/30 nm/80 nm of Cr/Pd/Au was deposited using electron-beam evaporation, followed by liftoff in acetone. The metal contacts to $WSe_2$ monolayer are placed at a distance of 1.5 µm away from the SiN waveguide, in order to reduce the propagation loss and minimize sheet resistance. A 10 nm/35 nm (100 loops/ 375 loops) layer of thermal ALD $Al_2O_3$ at 200 °/270 °C is then deposited to form the dielectric of the Gr-$Al_2O_3$-$WSe_2$ capacitor. In order to reduce the metal-$WSe_2$ contact resistance, we anneal the SiN waveguide with $Al_2O_3$ covered $WSe_2$ at 270 °C for 4 hours in vacuum. We then transfer and pattern monolayer graphene, as described in the section above, followed by vacuum annealing the composite Gr-$Al_2O_3$-$WSe_2$ on SiN waveguide at 275 °C for 4 hours in vacuum to remove PMMA residue left on graphene monolayer after the transfer and patterning. Following this, the metal contacts to the graphene layer is patterned using EBL and 5 nm/20 nm/50 nm of Cr/Pd/Au is then deposited using electron-beam evaporation, followed by liftoff in acetone. Similar to the metal placement configuration on WSe2 monolayers, the metal contacts are placed at an offset of 1.5 µm from the SiN waveguide. Finally, we define and wet etch (100:1 BOE) the vias to open the metal electrodes in contact with WSe$_2$, for testing. We achieve high electro-optic bandwidth in our devices by optimizing the graphene transfer process, involving multiple annealing steps in our device fabrication, depositing a 45 nm thick dielectric that minimizes pin-hole defects and finally optimizing the metal contacts to both the monolayer to reduce the contact resistance.

## Optical sheet conductivity of monolayer WSe$_2$ and graphene

We use the 2D sheet conductivity model to extract the electro-optic response of monolayer Gr and monolayer semiconductor WSe$_2$, as is commonly done when modelling graphene monolayers. The change in real part of the effective index of the composite SiN-2D waveguide ($\Delta n_{\text{eff}}$) in the top panel of figure 3a is a combination of the electro-refractive response of graphene i.e. $Im\left\{\frac{\sigma_G}{\sigma_0}\right\}$ and monolayer WSe$_2$ ($\Delta n_{\text{WSe2}}$). We predominantly attribute the change in the imaginary part of the effective index ($\Delta k_{\text{eff}}$) in the bottom panel of figure 3a, to the change in real part of the normalized complex conductivity i.e. $Re\left\{\frac{\sigma_G}{\sigma_0}\right\}$ of graphene with voltage. In accordance with the relation in equation (1), the $Re\left\{\frac{\sigma_G}{\sigma_0}\right\}$ is related to the imaginary part of dielectric permittivity, that contributes to absorption, whereas the $Im\left\{\frac{\sigma_G}{\sigma_0}\right\}$ is related to the real part of dielectric permittivity, that contributes to the change in index of monolayer graphene.

$$\sigma_G(\omega) = j\omega t_d \varepsilon_0 (\varepsilon(\omega) - 1) \tag{1}$$

Since the electro-optic response of graphene predominantly affects the $\Delta k_{\text{eff}}$, we extract the normalized sheet conductivity of monolayer graphene as a function of applied voltage, by comparing the measured $\Delta k_{\text{eff}}$ in our experiments to the simulated change obtained using COMSOL Multiphysics finite element model. We model the monolayer graphene as a conductive sheet, with surface charge density ($J = \sigma_G(\omega) \cdot E$), with conductivity given by equation (3). The optical properties of graphene can be tuned by doping graphene electrostatically[47], [48] i.e. by applying a voltage across the Gr-Al$_2$O$_3$-WSe$_2$ capacitor. The doping of graphene induces a shift in the fermi energy level of graphene ($E_F$), given by

$$E_F = \hbar v_F \sqrt{\left(\pi \left(\frac{\varepsilon_0 \varepsilon_R V}{de} + n_{initial}\right)\right)} \qquad (2)$$

where, $\varepsilon_0$ is the vacuum permittivity, $\varepsilon_R$ is the relative permittivity of the dielectric Al$_2$O$_3$ separating the two monolayers, $e$ is the electronic charge, $v_F$ is the fermi velocity in graphene, and $n_{initial}$ is the initial chemical doping of the graphene layer (which is dependent on the processing of graphene and on the substrate). The normalized optical conductivity of graphene ($\frac{\sigma_G}{\sigma_0}$) is related to the fermi level through the following equation

$$\frac{\sigma_G(\omega)}{\sigma_0} = \frac{1}{2}\left(\tanh\left(\frac{\hbar\omega+2E_F}{4k_BT}\right) + \tanh\left(\frac{\hbar\omega-2E_F}{4k_BT}\right) - \frac{i}{\pi}\left(\log\left(\frac{(\hbar\omega+2E_F)^2}{(\hbar\omega-2E_F)^2+(2k_BT)^2}\right)\right) + \frac{i8}{\pi}\left(\frac{E_F}{\hbar\omega+i\hbar\Upsilon}\right)\right) \qquad (3)$$

where $\sigma_0$ is the universal conductivity of graphene, $\hbar$ is the reduced Planck's constant, $\omega$ is the optical frequency, $k_B$ is the Boltzmann constant, T is the temperature and $\Upsilon$ is the intra-band carrier relaxation rate, assumed to be 100 fs, as predicted for similar structures. We find from our simulations that the graphene is initially p-doped with $(5.2 \pm 0.3) \times 10^{12}$ cm$^{-2}$ carriers (E$_{Finit}$ = 0.24 $\pm$ 0.006 eV)) and the slope of $\Delta k_{eff}$ indicates that the $\varepsilon_r = 6.9 \pm 0.2$. We model monolayer WSe$_2$, similar to graphene and is explained in detail in Ref [39]. We extract a change of $\sim$ 18% in the refractive index of monolayer WSe$_2$ with an electron doping density of $(2.54 \pm 0.74) \times 10^{13}$ cm$^{-2}$ at 30 V.